\documentclass[sigconf]{acmart}

\usepackage{times}
\usepackage{ctable}
\usepackage{multirow}
\usepackage{balance}
\usepackage{booktabs}
\usepackage{hyphsubst}
\usepackage{graphicx}
\usepackage{amsmath}
\usepackage{url}
\usepackage{breakurl}
\usepackage{hhline}
\usepackage[caption = false]{subfig}

\newcounter{lizcounter}
\DeclareRobustCommand{\liz}[1]{\textbf{/* #1 (liz) */}\stepcounter{lizcounter}\typeout{LaTeX Warning: liz comment \thelizcounter: #1 (line \the\inputlineno)}}

\newcounter{findingscounter}

\newcommand{\pred}{BLL$_{I,S}$}

\newcommand{\rand}{\textit{Random}}
\newcommand{\res}{\textit{CompSci}}

\newcommand{\para}[1]{\vspace{2mm}\noindent\textbf{#1}}

\renewcommand\footnotetextcopyrightpermission[1]{} %

\begin{document}

\title{Studying Confirmation Bias in Hashtag Usage on Twitter}

\author{Dominik Kowald}
\affiliation{%
  \institution{Know-Center GmbH}
  \city{Graz, Austria} 
}
\email{dkowald@know-center.at}
\author{Elisabeth Lex}
\affiliation{%
  \institution{Graz University of Technology \& Know-Center GmbH}
  \city{Graz, Austria} 
}
\email{elisabeth.lex@tugraz.at}


\begin{abstract}
The micro-blogging platform Twitter allows its nearly 320 million monthly active users to build a network of follower connections to other Twitter users (i.e., \textit{followees}) in order to subscribe to content posted by these users \cite{Myers2014,kwak2010}. With this feature, Twitter has become one of the most popular social networks on the Web and was also the first platform that offered the concept of \textit{hashtags} (see post by Chris Messina in 2007\footnote{\url{https://twitter.com/chrismessina/status/223115412}}). Hashtags are freely-chosen keywords, which start with the hash character ``\#'', to annotate, categorize and contextualize Twitter posts (i.e., tweets) \cite{romero2011,Huang2010}.

\para{Problem.}
Although hashtags are widely accepted and used by the Twitter community, the heavy reuse of hashtags that are popular in the personal Twitter networks (i.e., own hashtags and hashtags used by followees) can lead to filter bubble effects \cite{pariser2011filter} and thus, to situations, in which only content associated with these hashtags are presented to the user. These filter bubble effects are also highly associated with the concept of confirmation bias, which is the tendency to favor and reuse information that confirms personal preferences \cite{plous1993psychology}. One example would be a Twitter user who is interested in political tweets of US president Donald Trump. Depending on the hashtags used, the user could either be stuck in a pro-Trump (e.g., ``\#MAGA'') or contra-Trump (e.g., ``\#fakepresident'') filter bubble.

Therefore, the goal of this paper is to study confirmation bias and filter bubble effects in hashtag usage on Twitter by treating the reuse of hashtags as a phenomenon that fosters confirmation bias.

\para{Method.}
To reach this goal, we crawl two datasets from Twitter. The first one (i.e., \res{} dataset) consists of researchers from the field of computer science (= seed users) and their followees, while the second one (i.e., \rand{} dataset) consists of random people (= seed users) and their followees. The statistics of these datasets are depicted in Table \ref{tab:datasets} and details of the crawling strategy are described in \cite{kowald2017temporal}. For all the seed users (i.e., $|U_S|$) in these datasets, we analyze (i) individual hashtag reuse (i.e., reusing own hashtags), and (ii) social hashtag reuse (i.e., reusing hashtags, which has been previously used by a followee) with respect to hashtag usage types and temporal effects.

Finally, we discuss ways to overcome confirmation bias and filter bubble effects in Twitter by means of recommendation mechanisms that focus on beyond-accuracy metrics.

\para{Code.} For reasons of reproducibility, we conduct this study using our open-source tag recommendation and benchmarking framework \textit{TagRec} \cite{kowald2017tagrec,trattner2015tagrec}, which is available via our GitHub repository\footnote{\url{https://github.com/learning-layers/TagRec}}.

\begin{table}[t!]
	\small
  \setlength{\tabcolsep}{3.8pt}	
  \centering
    \begin{tabular}{l||cccccc}
    \specialrule{.2em}{.1em}{.1em}
											Dataset				& $|U_S|$			& $|U|$				& $|T|$					& $|HT|$			& $|HTAS|$				\\\hline 
											\res{}				&	2,551				&	91,776			&	5,649,359			&	1,081,403		&	9,161,842					\\\hline
											\rand{}				& 3,466		  	& 127,112 		& 8,157,702 		& 1,507,773		& 13,628,750				\\
		\specialrule{.2em}{.1em}{.1em}								
    \end{tabular}
    \caption{Statistics of our \res{} and \rand{} Twitter datasets. Here, $|U_S|$ is the number of seed users, $|U|$ is the number of distinct users, $|T|$ is the number of tweets, $|HT|$ is the number of distinct hashtags and $|HTAS|$ is the number of hashtag assignments. We aim to study confirmation bias in hashtag usage on Twitter by analyzing (i) the reuse rate of these hashtag assignments, and (ii) temporal effects of hashtag usage.
\vspace{-6mm}}	
  \label{tab:datasets}
\end{table}

\begin{figure}[t!]
   \centering
      \includegraphics[width=0.48\textwidth]{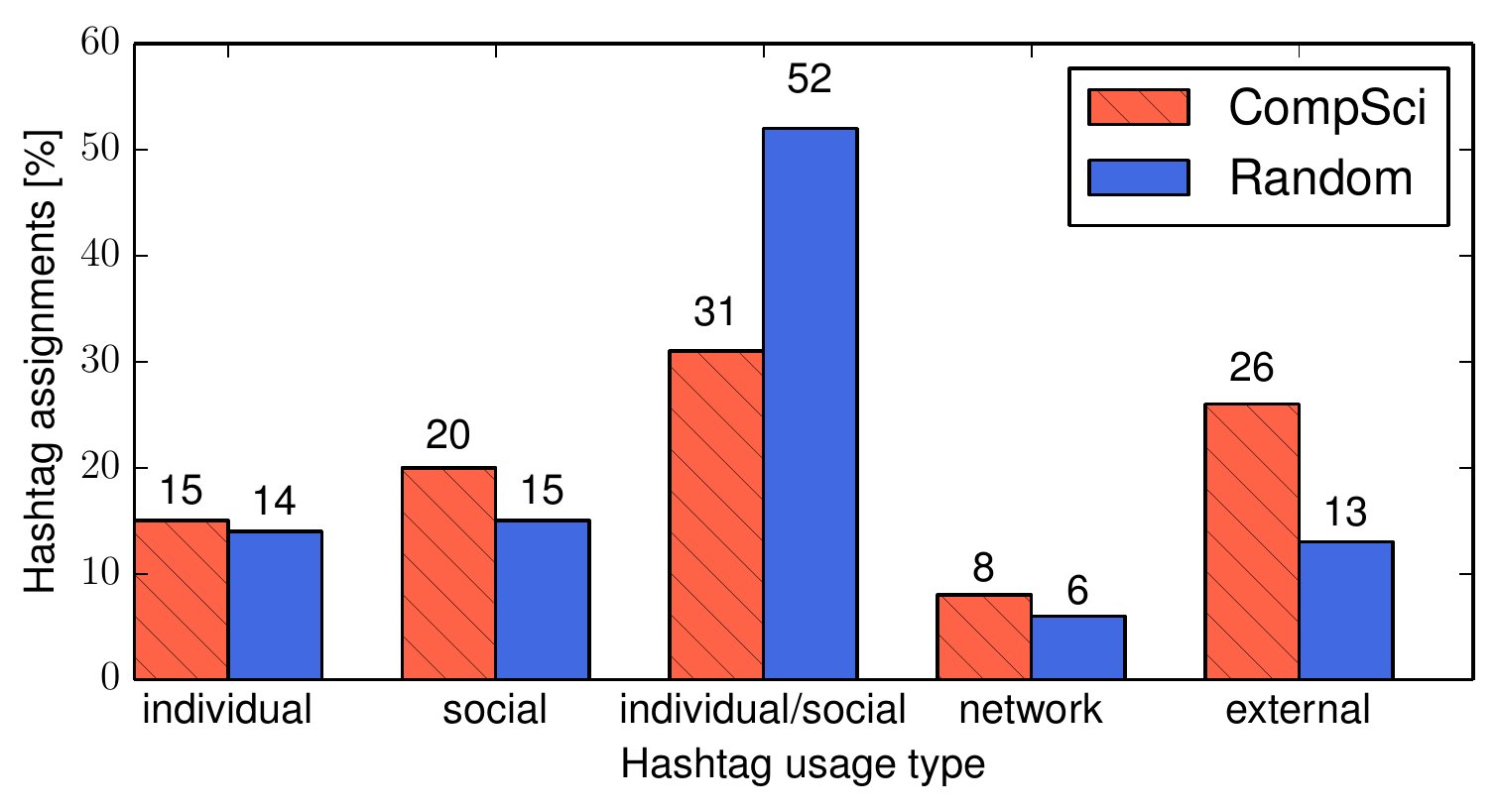} 
   \caption{Analysis of hashtag reuse types as a proxy for studying confirmation bias. For each hashtag assignment, we study whether the corresponding hashtag has been used by the same user before in time (``individual''), by some of the followed users (``social''), by both (``individual/social''), by anyone else (``network'') or neither of them (``external''). We find that between 66\% and 81\% of hashtag assignments in both datasets can be explained by individual or social hashtag usage (i.e., the sum of ``individual'', ``social'' and ``individual/social''). This is a clear indication of confirmation bias in hashtag usage on Twitter.
\vspace{-3mm}}
	 \label{fig:intro}
\end{figure}

\para{Results.}
In our datasets, we first analyze hashtag assignments as well as hashtag reuse practices with the aim of studying confirmation bias in hashtag usage on Twitter. Specifically, for each hashtag assignment, we study whether the corresponding hashtag has either been used by the same user before (``individual''), by some of her followees (``social''), by both (``individual/social''), by anyone else in the dataset (``network'') or by neither of them (``external''). The results of this study are shown in Figure \ref{fig:intro}. We find that 66\% of hashtag assignments in the \res{} dataset and 81\% in the \rand{} dataset can be explained by individual or social hashtag reuse, which is a clear indication of confirmation bias and filter bubble effects.

\begin{figure*}[t!]
   \centering
	 \captionsetup[subfigure]{justification=centering}
	 \subfloat[][Individual hashtag reuse\\\res{} dataset]{ 
      \includegraphics[width=0.24\textwidth]{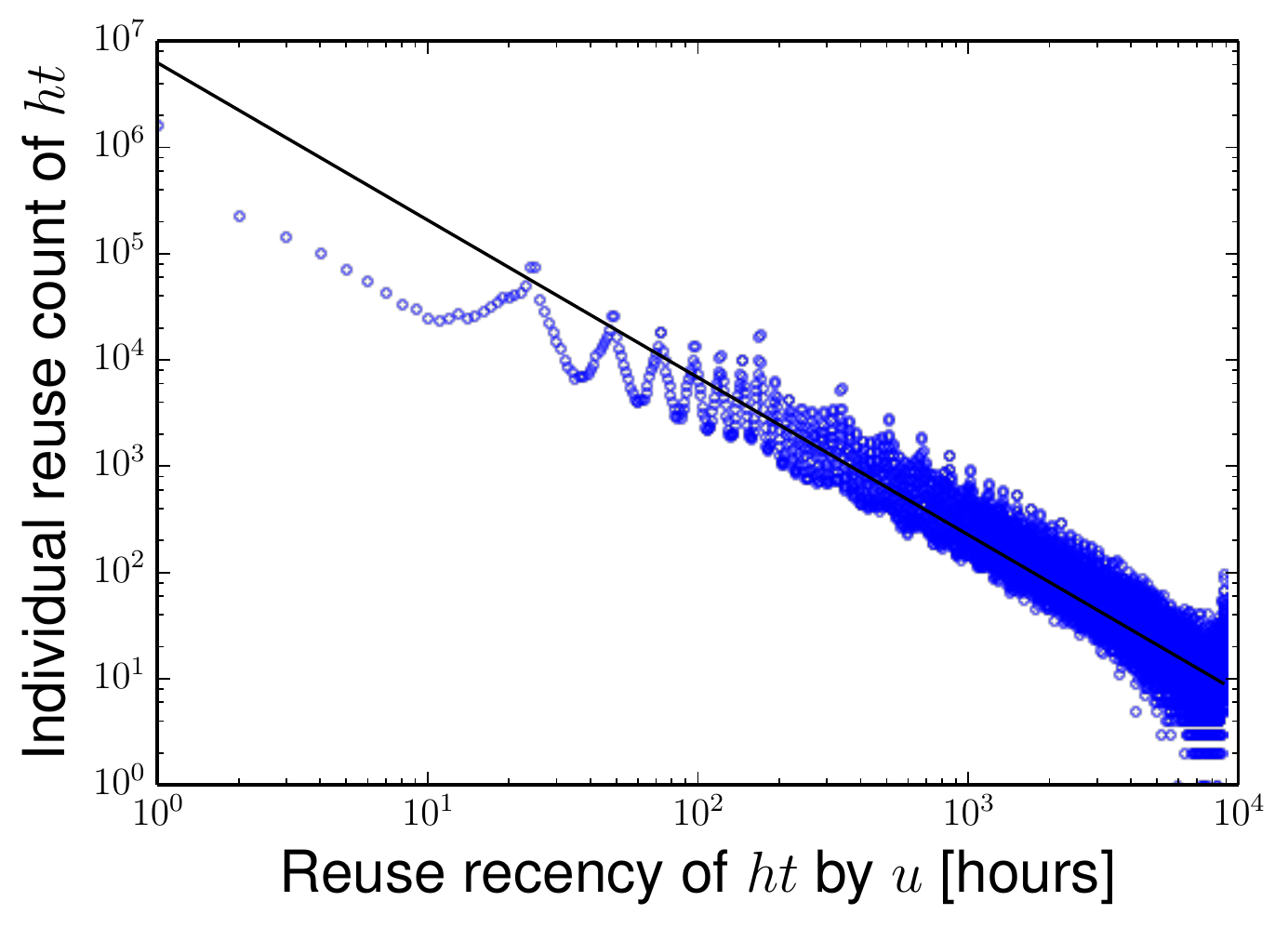} 
   }
	 \subfloat[][Individual hashtag reuse\\\rand{} dataset]{ 
      \includegraphics[width=0.24\textwidth]{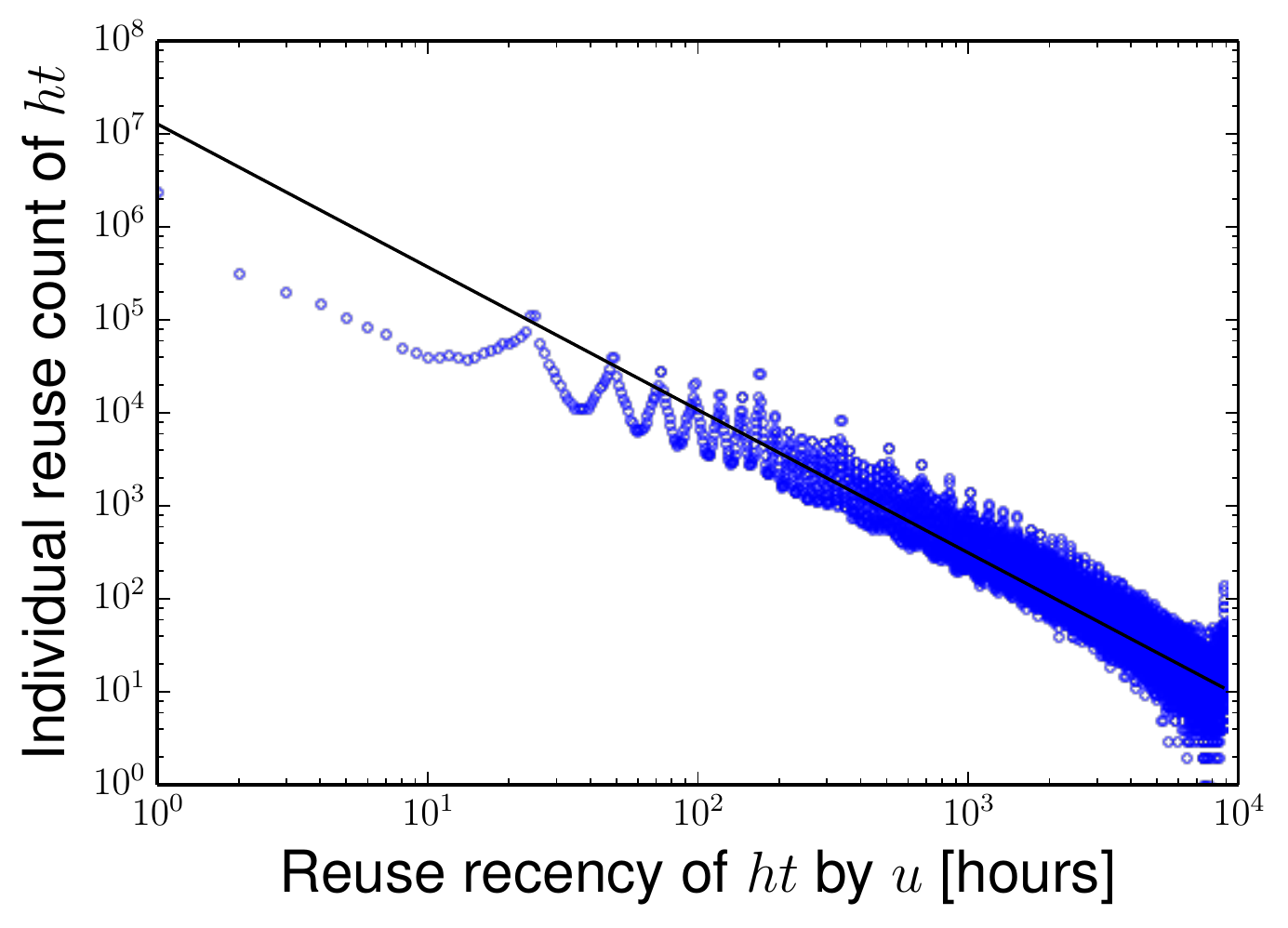} 
   }
	 \subfloat[][Social hashtag reuse\\\res{} dataset]{ 
      \includegraphics[width=0.24\textwidth]{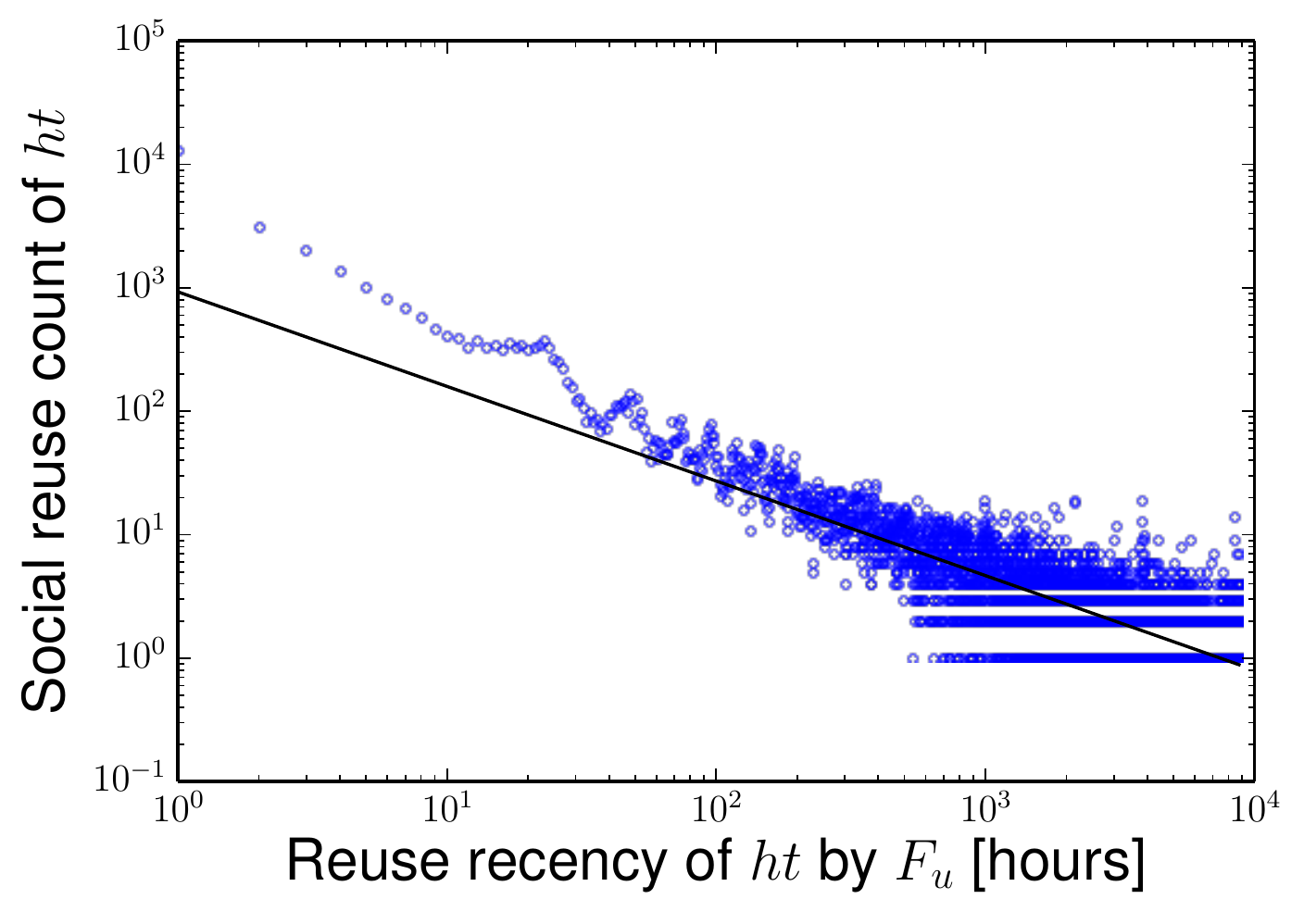} 
   }  
	 \subfloat[][Social hashtag reuse\\\rand{} dataset]{ 
      \includegraphics[width=0.24\textwidth]{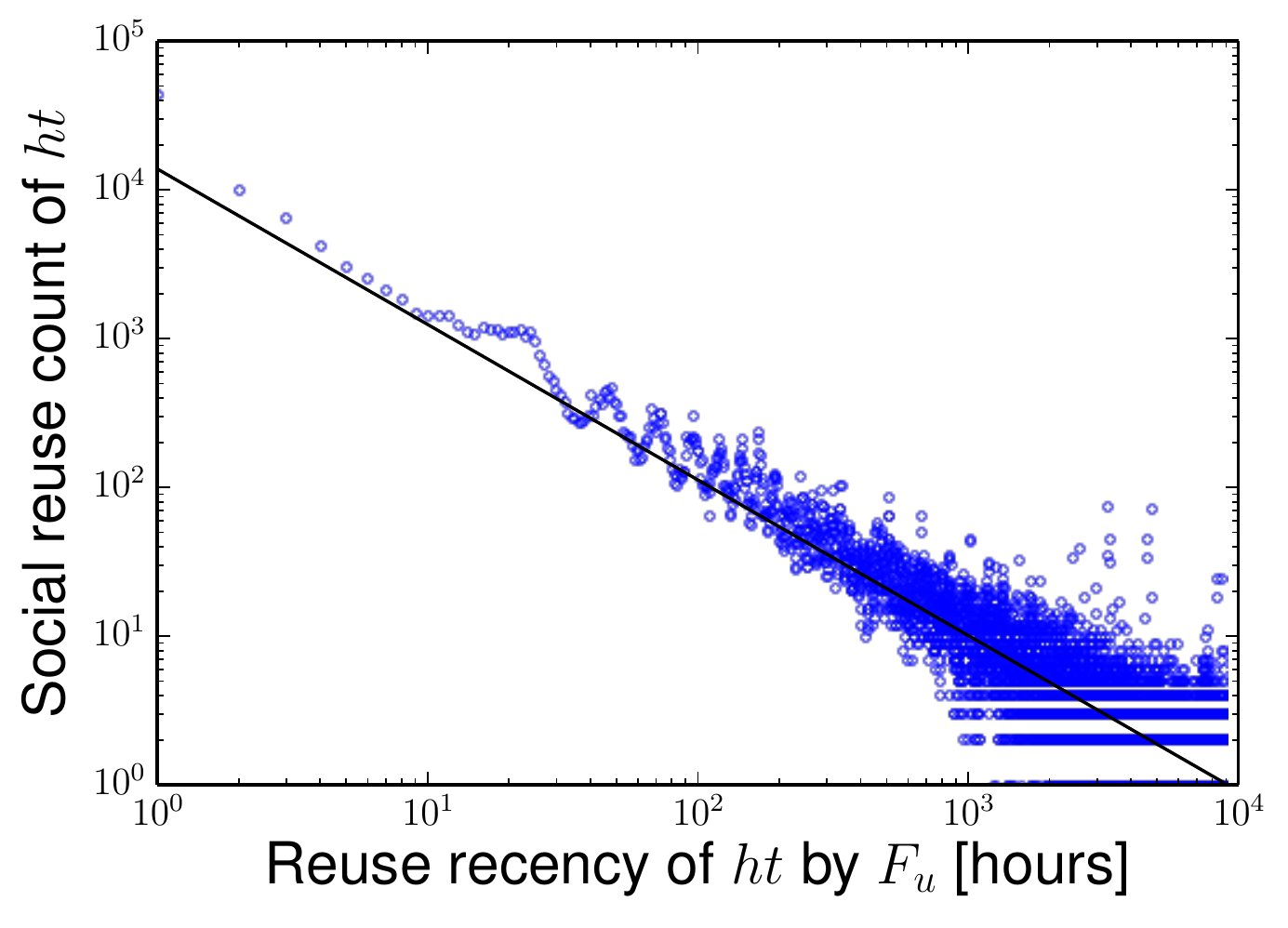} 
   }
   \caption{Temporal effects on individual and social hashtag reuse for the \res{} and \rand{} datasets (plots are in log-log scale). Plots (a) and (b) show that the more recently a hashtag $ht$ was used by a user $u$, the higher its individual reuse count (i.e., people tend to reuse hashtags that have been used very recently by their own). Plots (c) and (d) show that the more recently a user $u$ was exposed to a hashtag $ht$, which was used by the followees $F_u$, the higher its social reuse count (i.e., people tend to reuse hashtags that have been used recently in the social network). We find that temporal effects play an important role in individual and social hashtag reuse, which is our measure for studying the confirmation bias in hashtag usage on Twitter.
\vspace{-3mm}}
	 \label{fig:analysis}
\end{figure*}

Next, we study to what extent temporal effects play a role in the reuse of individual and social hashtags on Twitter. Specifically, we analyze the recency of hashtags assignments (i.e., the time since the last hashtag usage). The results for this study are shown in the plots of Figure \ref{fig:analysis} and we find that temporal effects have an important influence on both individual as well as social hashtag reuse. People tend to reuse hashtags that were used very recently by their own and/or by their Twitter followees. Interestingly, there is a clear peak after 24 hours in all four plots, which further indicates that users typically (re)use the same set of hashtags in this time span and thus, tend to tweet about similar topics on a daily basis. This shows that also temporal effects are an important influence for confirmation bias and filtering bubble effects in hashtag usage on Twitter.

\para{Conclusion \& Discussion.} Our findings of this paper are two-fold:

\begin{enumerate}
\item We find that between 66\% and 81\% of hashtag assignments in our datasets can be explained by individual or social hashtag reuse, which is a clear indication of confirmation bias in hashtag usage on Twitter.
\item We find that people tend to reuse hashtags that were used very recently by their own and/or by their Twitter followees, which means that confirmation bias in hashtag usage  is also strongly influenced by temporal effects.
\end{enumerate}

Based on this, we now discuss ways to overcome confirmation bias and filter bubble effects by using recommendation mechanisms. In our previous work (i.e., \cite{kowald2017temporal}), we presented a novel hashtag recommendation algorithm, which utilizes the Base-Level Learning (BLL) equation of the cognitive architecture ACT-R \cite{anderson2004integrated} for suggesting hashtags to users. Specifically, this algorithm ranks the hashtags of a user and her followees based on the factors of frequency and recency (see also \cite{Kowald2016a,springer_bllac}). Although this approach achieves good results with respect to recommendation accuracy (see \pred{} in Figure \ref{fig:results}), it fosters hashtag reuse and thus, confirmation bias and filter bubbles effects in our two Twitter datasets.

\begin{figure}[t!]
   \centering
	 \captionsetup[subfigure]{justification=centering}
	 \subfloat[\res{} dataset]{ 
      \includegraphics[width=0.24\textwidth]{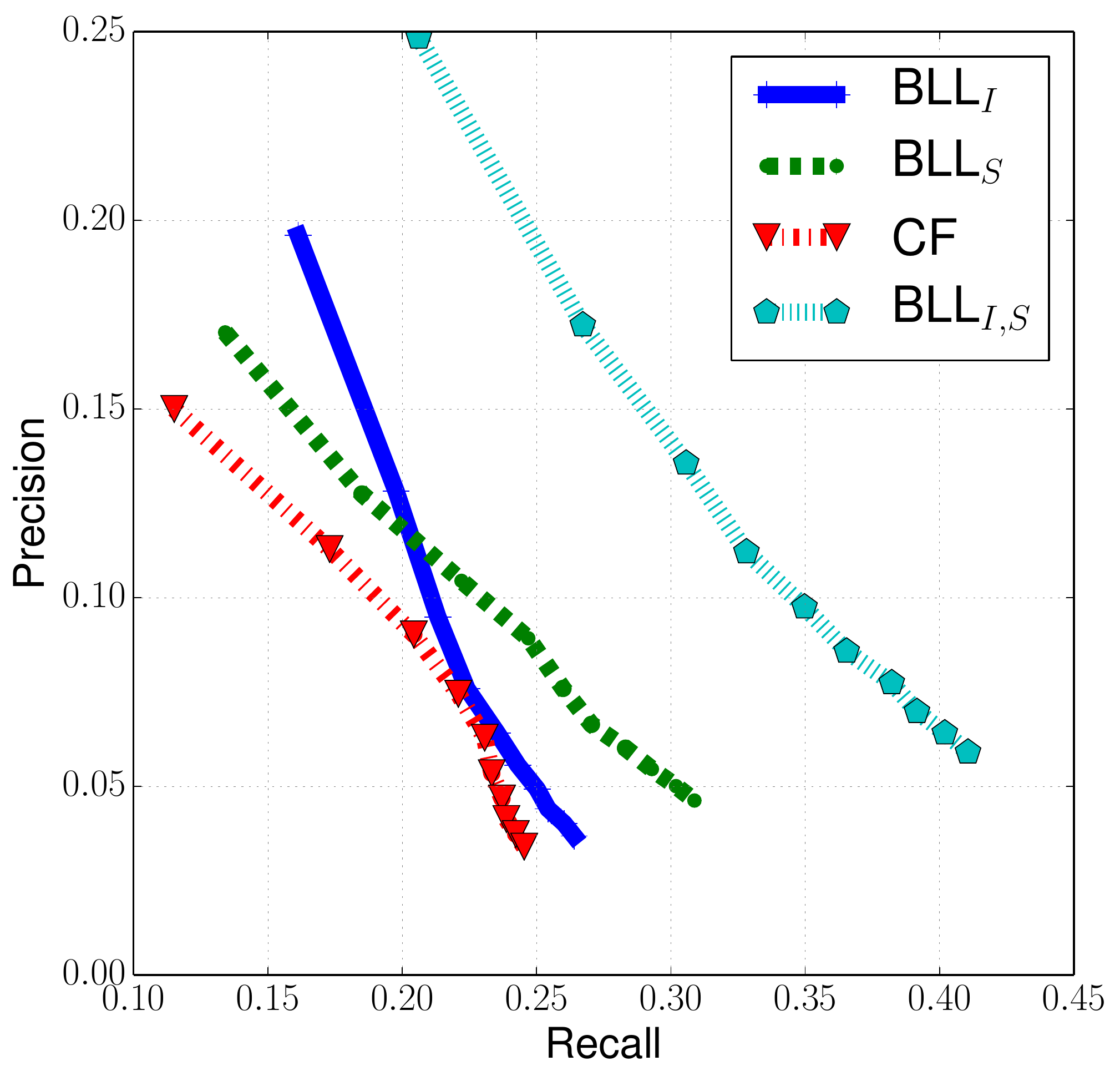} 
   }
	 \subfloat[\rand{} dataset]{ 
      \includegraphics[width=0.24\textwidth]{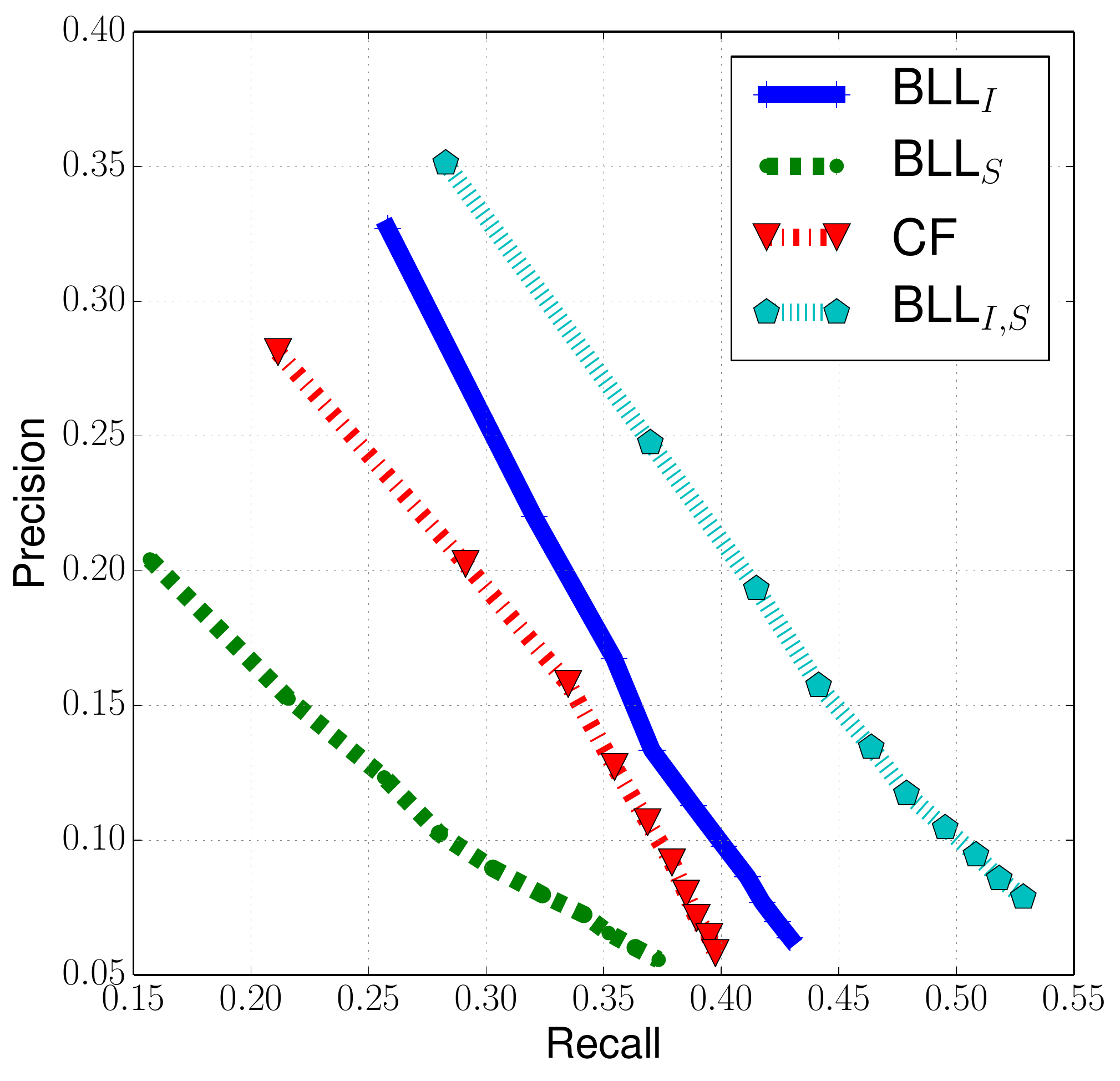} 
   }
   \caption{Precision / Recall plots showing the accuracy of BLL$_I$ (i.e., only individual hashtags), BLL$_S$ (i.e., only social hashtags), CF (i.e., Collaborative Filtering), \pred{} (i.e., individual and social hashtags) for $k$ = 1 - 10 recommended hashtags. \pred{} provides the best results with respect to predication accuracy but also fosters the reuse of hashtags and thus, confirmation bias and filter bubbles effects in our two Twitter datasets.
\vspace{-3mm}}
	 \label{fig:results}
\end{figure}

In order to overcome these effects, we propose to focus on beyond-accuracy metrics of recommender systems such as diversity and serendipity \cite{kaminskas2017diversity,Kowald2015}. If we again take the example of a Twitter user who is interested in political tweets of US president Donald Trump, we could think of a hybrid recommendation approach with the goal to achieve a trade-off between accuracy, serendipity and diversity. For example, if we think of a pro-Trump Twitter user, a typical accuracy-focused recommender system would only recommend pro-Trump tweets and hashtags (e.g., ``\#MAGA''). Thus, our idea is to mix these pro-Trump recommendations also with contra-Trump recommendations (e.g., ``\#fakepresident'') to boost serendipity and diversity, and to give the user the chance to break the filter bubble.

\para{Future Work.} Finally, for future work, we also plan to verify our findings in larger Twitter data samples than the ones used in this paper as well as in other online social networks that feature the concept of hashtags, such as Instagram and Facebook.

Additionally, it would be interesting to evaluate the influence of social hashtags exposure, for example, by investigating reweet or mention networks in Twitter.

\para{Acknowledgments.} This work is funded by the Know-Center GmbH, which is part of the Austrian COMET program, and the H2020 project AFEL under grant agreement 687916.

\para{Keywords.} Twitter; Hashtag Usage; Confirmation Bias; Filter Bubble; Tag Recommendations; ACT-R; TagRec

\end{abstract}

\maketitle

\balance

\end{document}